\documentclass[a4paper,twocolumn]{esapub_ac}
\usepackage{fancybox}

\usepackage{color,pstcol}
\usepackage{pst-grad}
\usepackage{graphicx}
\usepackage{epsf}
\usepackage{times}
\usepackage{natbib}
\def\sun{\hbox{$\odot$}}

\title{A radial velocity search for p-modes in $\beta$\,Vir}
\author{M.~Marti\'{c}${^1}$, J.C.~Lebrun${^1}$, T.~Appourchaux${^2}$, J.~Schmitt${^3}$} 

\affil{${^1}$Service d'A\'eronomie du CNRS, BP No 3, 91371
   Verri\`eres le Buisson, France\\${^2}$Institut d'Astrophysique Spatiale, Universit\'e Paris XI - CNRS,
F-91405 Orsay Cedex, France\\${^3}$O.H.P. 04870 St. Michel l'Observatoire, France}

\begin{document} 
\maketitle


\begin{abstract}

Spectroscopic high-resolution observations were performed
with fiber-fed
 cross-dispersed echelle spectrographs in order to measure
 the fluctuations in radial velocities of a sample of bright stars
 that are likely
 to undergo solar-like oscillations.
Here we report the results for $\beta$\,Vir (HR4540) from two observing runs
carried out in February 2002 with FEROS at the ESO 1.52
 m telescope in La Silla (Chile) and ELODIE spectrograph at 1.93 OHP telescope
 (Observatoire de Haute Provence, France). The analysis of the time series
 of Doppler shifts from both sites has revealed the presence of
 an excess power around 1.7\,mHz.
We discuss the interpretation  of this data set in terms of possible
$p$-mode oscillations.

\end{abstract}

\section{\textsc{Introduction}}
%\section*{Introduction}

Precise measurements of the frequencies and amplitudes of $p$-mode oscillations for solar-like stars lead to detailed inferences about the internal structure and provide the most powerful constraint to the theory of the stellar evolution \citep[see the review by][]{Christ02}. Stellar fundamental parameters, most notably mass, are usually well specified in binary pairs. $\beta$\,Vir binary system is an exception, it offers a challenging task to asteroseismology to discriminate between possible mass estimations from stellar models and astrometric measurements. The so-called mean large frequency separation $\Delta\nu_{0}$
between $p$-modes of same degree and adjacent $n$, reflects the stellar density and scales approximately as $\Delta\nu_{\rm 0 }\propto\,(M/R^{3})^{1/2}$. The stellar 
basic parameters and expected values for $\Delta\nu_{\rm 0}$ of F9V  $\beta$\,Vir (HR4540, HD 102870, HIP 57757, M$_{\rm V}$=3.61) are :

%tab-charbvir.tex
% target stars characteristics
%\begin{tabular}{llllllllll}
\begin{tabular}{clllllllll}
            \hline
            \noalign{\smallskip}
  %HR   &   $\Pi$(mas) & V & T$_{eff}$(K) & L/\Lsun & log\,$g$ & M/\Msun & $\nu_{\rm max}$ & $\Delta\nu_{\rm 0}(\mu{\rm Hz})$  \\
    $\Pi$ &  $T_{\rm eff}$ & $L/L_{\sun}$ & log\,$g$ &  $M/M_{\sun}$ & $\Delta\nu_{\rm 0 }$ \\
     (mas) & (K) &  &  &   &   ($\mu{\rm Hz})$ \\
            \noalign{\smallskip}
            \hline
91.74 &  6109 & 3.57 & 4.20  & 1.35-1.6 & 71-78 \\ 
            \noalign{\smallskip}
            \hline
         \end{tabular}

The observations of $\beta$\,Vir were carried out over 12 nights in
 February 2002. The run started on February 18 at OHP and finished on
 February 28. The observations of the same star with FEROS at La Silla Observatory were made over five nights (February 24 - March 1). 
Because of the
 high pressure of the observing requests for both instruments
 it was difficult to obtain the optimum period for the observation of this star. Note that the number of possible observable targets at both sites
 is extremely limited. The candidate star for the detection of
 the p-mode oscillations by spectroscopic method needs
 to have a solar-type rich spectrum of narrow lines
 and to have a high S/N ratio in order to attain photon noise
 limit in power spectra. Here, we report the clear detection, independently
with two instruments, of excess power, providing evidence for solar-like oscillations in $\beta$\,Vir.

%\section{\textsc{Instrumentation}}
%\section*{Instrumentation}

%\input{Instr_ohp.tex}

%\input{Instr_feros.tex}

%------------------------------------------------------------------------------
\section{\textsc{Data reduction}}
%\section*{Data reduction}

% Fig_specohpferord.tex
   \begin{figure}
\vspace{0cm}
\resizebox{\hsize}{!}{\includegraphics
%{/spectro/feros2002/esaWP1/rap_esa_martic/spectre_feros-elodie_r.eps}}
{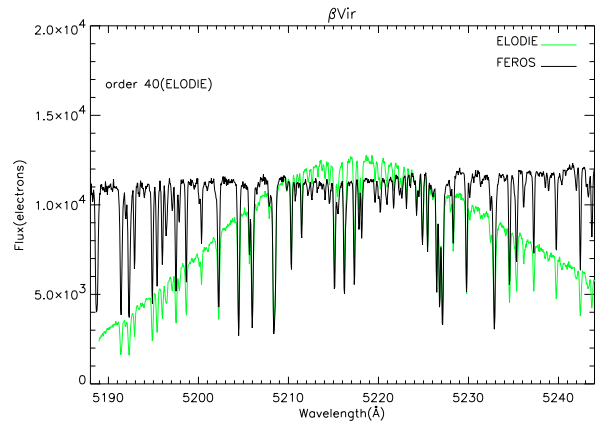}}

\vspace{0cm}
      \caption[]{Example of the ELODIE and FEROS spectra
 (scaled intensity) in the same $\delta\lambda$ region. Typical exposure time was 90\,s with ELODIE/1.93m and 40\,s with FEROS/1.52m.
}
         \label{Fig_specohpferord}
   \end{figure}

% Fig_feros1dspectrum.tex
   \begin{figure}
\vspace{0cm}
\resizebox{\hsize}{!}{\includegraphics
%{/spectro/feros2002/esaWP1/rap_esa_martic/spectre_bvir1.eps}}
{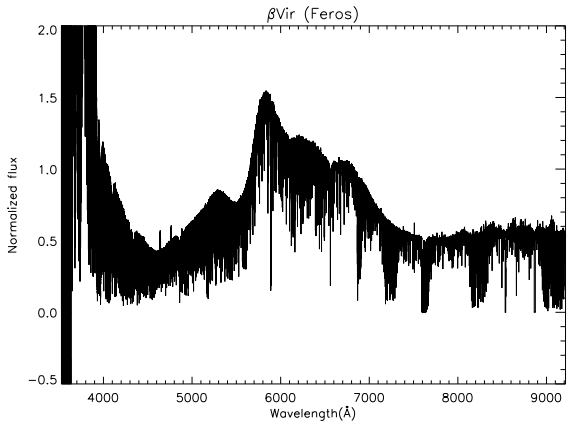}}
\vspace{0cm}
      \caption[]{FEROS merged 1-D spectrum. The spectrum is not "flat" partly because of the energy distribution of the object (modified by the earth atmosphere), but dominantly by the bimodal intensity distribution of the flat-field spectra.
}
         \label{Fig_feros1dspectrum}
   \end{figure}

The observations with ELODIE are now largely automated, the raw frames are reduced in real time and "only" 2-D extracted order-pixels (1024x67) images are saved. The instrument and the reduction process are explained in Martic et al. (1999). Here we show several results from FEROS reduction steps since this instrument was used for the first time for the detection of the solar-like oscillations.
The observational set-up configuration was: 
\begin{itemize}
\item Telescope : ESO 1.52-m telescope at La Silla
 \item Spectrograph :  bench-mounted echelle spectrograph (FEROS) with the fibres located at the Cassegrain focus
 \item Wavelength range : 39 orders from 360 to 920\,nm
 \item Resolution : $\sim$\,48000
 \item Detector : thinned and back-illuminated 2kx4k EEV CCD with its excellent quantum efficiency (QE = 98\,\% at 450\,nm) contributes considerably to the high efficiency of FEROS
 \item Read-out noise : 3.7 and 3.5\,e$^{-}$/pixel for channel A and B
 \item ThAr comparison lamp simultaneously with the star exposure: monitor the spectrograph variations
\end{itemize}

At the first site (OHP), we used larger telescope but we got lower average S/N ratio due to seeing conditions and overall instrument efficiency (S/Nmax$\sim$\,150 with ELODIE comparing to S/Nmax$\sim$\,300 with FEROS).
Note however that several serious problems and errors were found in course of the reduction of the data with ESO/FEROS pipe-line. We corrected as much as possible the evident errors but full development was out of scope of the present work. A new pipe-line (different from the one for OHP data) was developed to calculate the reference spectra, the mask spectra, quality factors etc. and to test the computation of the Doppler shifts from 2-D (39x4102) order-pixels images (see e.g one extracted order in Fig.~\ref{Fig_specohpferord}) or/and from 1-D merged calibrated spectra (see Fig.~\ref{Fig_feros1dspectrum}).

%---------------------------------------------------------------------------

\begin{figure}
\vspace{0cm}
\resizebox{\hsize}{!}{\includegraphics
%{/spectro/feros2002/esaWP1/rap_esa_martic/spectres_B-bvir.eps}}
{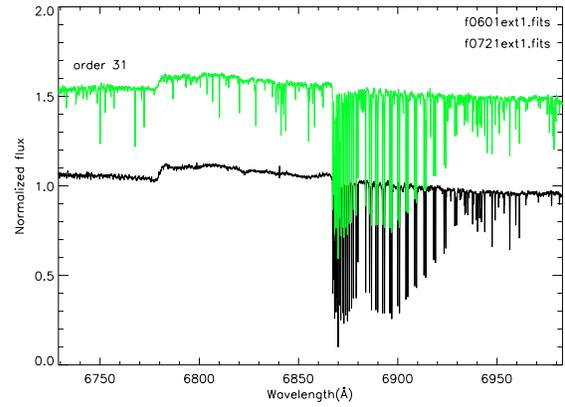}}
\vspace{0cm}
      \caption[]{$\beta$\,Vir spectrum (in green) superposed to the one from HR2142 in the spectral region contaminated by the telluric lines.}
         \label{Fig_spectresBbvir}
   \end{figure}

At the beginning of each night, we observed also a B star for an another ESO program. The spectrum of a B star allowed to produce a good mask for the contaminated spectral range and also to locate the slopes or unusual gradients over the extracted orders. Fig.~\ref{Fig_spectresBbvir} shows a $\beta$\,Vir spectrum (in green) superposed to the one from HR2142 in the spectral region contaminated by the telluric lines.

% Fig_spectresBbvir.tex

% Fig_derivefibre.tex
   \begin{figure}
\vspace{0cm}
\resizebox{\hsize}{!}{\includegraphics
%{/spectro/feros2002/esaWP1/THferos25.eps}}
{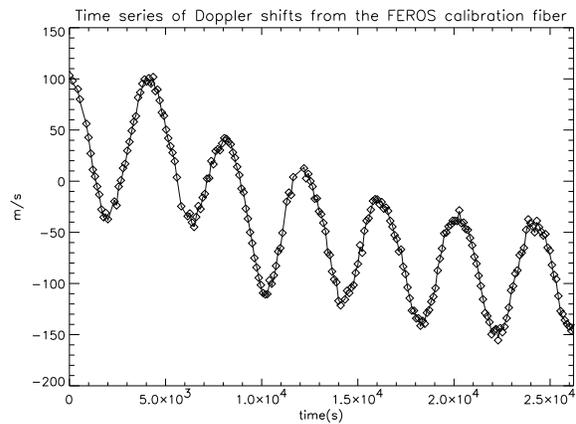}}
\vspace{0cm}
      \caption[]{Curve of the FEROS spectrograph motion on February 28, 2002. The extracted spectra from the second fibre (ThAr) were used to determine the instrument stability during the night.}
         \label{Fig_derivefibre}
   \end{figure}

In order to obtain the high precision in determination of spectral line displacements we proceed in the following manner: At first, the guess value for the Doppler shift is obtained from the comparison of the 1-D merged spectrum with the reference one. The current spectrum is then numerically shifted for this guess value. Note that we stretch spectra instead of simple shift. Residual small shifts are then calculated for each order from 2-D frames. The method used for Doppler shift calculations is optimum for small velocity displacement and gives better precision than the correlation technique. We have discovered that the FEROS instrument had large oscillatory motion during the night (see Fig.~\ref{Fig_derivefibre}). The amplitude of the oscillation from one night to another varies from 100 to 150\,m/s with a period around one hour. The main reason is probably due to imperfect temperature regulation. Fortunately the frequency of the FEROS temperature regulation/motion is well below the star oscillatory signature.
%---------------------------------------------------------------------------

\section{\textsc{Radial velocities}}

% Fig_obsohpferos.tex
   \begin{figure}
\vspace{0cm}

\resizebox{\hsize}{!}{\includegraphics
%{/export/home//milena/song04/obs_elodie-afoe.eps}}
{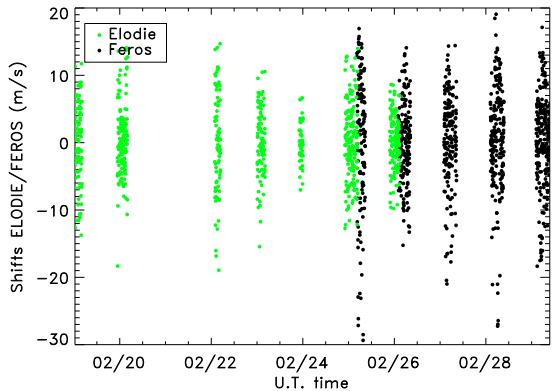}}
\vspace{0cm}
      \caption[]{
Doppler shift measurements over 10 nights from two-sites observation of $\beta$\,Vir : Elodie (in green), FEROS (in black). .
}
         \label{Fig_obsohpferos}
   \end{figure}

%\resizebox{\hsize}{!}{\includegraphics
%{/export/home//milena/song04/obs_elodie-afoe.eps}}

$\beta$\,Vir was observed over a campaign of twelve nights (2002, February 18 - 1 March). Telescope time accorded by the scientific committees allowed to have four nights of co-ordinated measurements at OHP and La Silla Observatory; because of the weather condition at OHP only two nights overlap. In Fig.~\ref{Fig_obsohpferos}, we display the resulting velocity measurements over 10 nights with ELODIE and FEROS. For each night and instrument, radial velocities were computed relative to the reference spectra.
  
% Fig_ts25ohpferos.tex
   \begin{figure}
\vspace{0cm}
\resizebox{\hsize}{!}{\includegraphics
%{/export/home//milena/song04/bvirohpferos2602.eps}}
{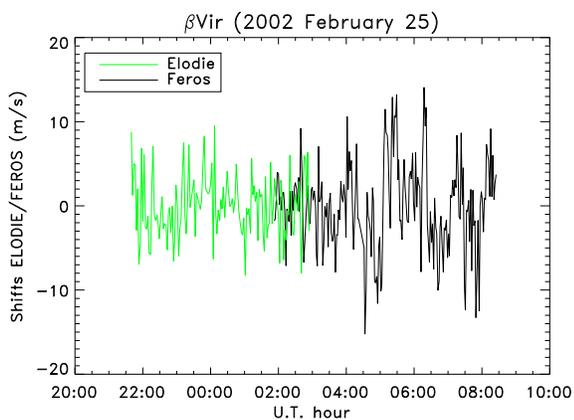}}
\vspace{0cm}
      \caption[]{Joint observations over one night: Elodie (in green), FEROS (in black).
}
         \label{Fig_ts25ohpferos}
   \end{figure}

%resizebox{4.6in}{!}{\includegraphics{bvirohpferos2602.eps}}
 
In Fig.~\ref{Fig_ts25ohpferos}, we show the dispersion of joint observations ELODIE/FEROS over one night. The typical rms for one night with ELODIE is of the order of 4\,m/s. The higher dispersion of the FEROS measurements is due to the errors of a new guiding system (for a bright star the centre-of gravity method is used).

%---------------------------------------------------------------------------
\section{\textsc{Power spectra analysis}}

For the power spectra analysis, different velocity time series were constructed,
 depending on the
window function and data quality \citep{Martic02}. The weights are assigned to each data point according to its estimated uncertainty in the velocity measurement. Here we present the results obtained using only local weights, for the computation of the power spectra, and without weights, for the computation of the Cleaned spectra \citep{Roberts87}. One can change significantly the noise/alias contribution \citep{Butler04}
by adjusting the weights for the data obtained with different precision from two instruments. This optimization process for the weights (which degrades the window function) has still to be tested.

% Fig_fftohpferos.tex
   \begin{figure}
\vspace{0cm}
\resizebox{\hsize}{!}{\includegraphics
%{/export/home//milena/song04/fft.eps}}
{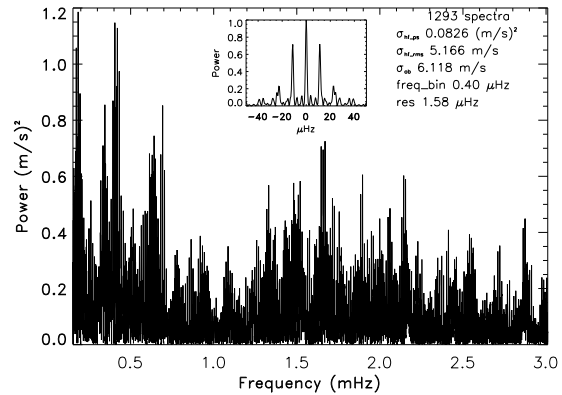}}

\vspace{0cm}
      \caption[]{Power spectrum of the Doppler shift measurements, computed with 0.4\,$\mu$Hz resolution, from joint ELODIE/FEROS observations. The inset shows the power spectrum of the window function for a signal amplitude of 1\,m/s and the same sampling rate as the observations.
}
         \label{Fig_fftohpferos}
   \end{figure}

%\resizebox{7in}{!}{\includegraphics{fft.eps}}
%\resizebox{7in}{!}{\includegraphics{fftbvirohp22-26feros26-1wbed.eps}}

In Fig.~\ref{Fig_fftohpferos}, we show the power spectrum of the Doppler shift measurements from joint ELODIE/FEROS observations. The data are not filtered at low frequencies. The excess power between 1.2 and 2.3\,mHz, is the signature of solar-like oscillations in $\beta$\,Vir.

%---------------------------------------------------------------------------

% Fig_fftsepohpferos.tex
   \begin{figure}
\vspace{0cm}
\resizebox{\hsize}{!}{\includegraphics
%{/export/home//milena/song04/powerbvirohpferossep.eps}}
{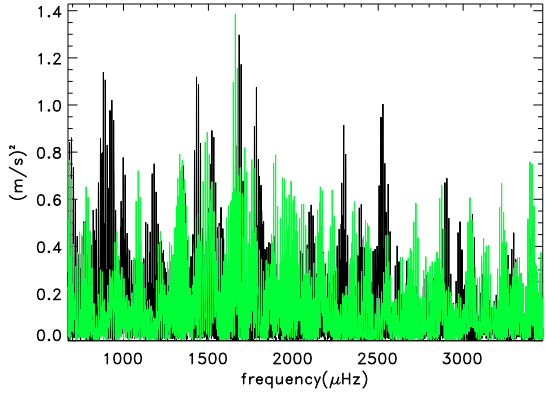}}

\vspace{0cm}
      \caption[]{Comparison of the power spectra from Elodie (in green) and FEROS (in black)
observing runs. 
}
         \label{Fig_fftsepohpferos}
   \end{figure}

%\resizebox{7in}{!}{\includegraphics{powerbvirohpferossep.eps}}

By comparing the power spectra from Elodie and FEROS separate
observing runs (see Fig.~\ref{Fig_fftsepohpferos}) one can see that the hump of excess power consistent with oscillations, centred at 
about 1700\,$\mu{\rm Hz}$ is present in both spectra. This confirms the stellar origin of the excess power detected independently with two instruments on two different sites.

\begin{figure}
\vspace{0cm}
\resizebox{\hsize}{!}{\includegraphics
%{/export/home//milena/song04/combvir85.eps}}
{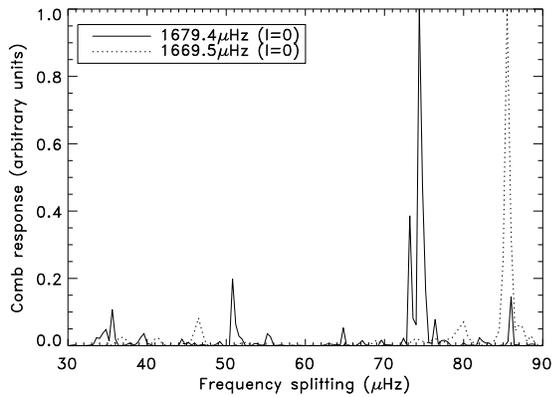}}
\vspace{0cm}
      \caption[]{Comb responses computed at two central frequencies of the power and cleaned spectra of the combined velocity time series.}
         \label{Fig_combvir}
   \end{figure}

To estimate the large frequency separation $\Delta\nu_{0}$ between $p$-modes of same degree and adjacent $n$ in the region of excess power we used the comb response method \citep[see][]{Kjeldsen95}. Because of 
the interaction of modes
 with different phases, daily gaps and noise which suppress and/or reinforce peaks by mixing multiple frequencies in power spectra \citep[see][]{Martic04}, we obtained
the maximum Comb response for the two possible values of the large frequency spacing:
 $\Delta\nu_{\rm 0 }$=85.5\,$\mu$Hz from the power spectrum
 and $\Delta\nu_{\rm 0 }$=74.5\,$\mu$Hz from the corresponding
 Cleaned spectrum of the same time sequence.
 Comb responses (CR) computed at two central frequencies of the power and cleaned spectra time series are shown in Fig.~\ref{Fig_combvir}. The maximum CR from the power spectrum is probably shifted (dashed line) because of more important contribution of the aliases.

%\resizebox{4.6in}{!}{\includegraphics{sfsbvir.eps}} %
% Fig_combvir.tex

\begin{figure}
\vspace{0cm}
\resizebox{\hsize}{!}{\includegraphics
%{/export/home//milena/song04/sfsbvir.eps}}
{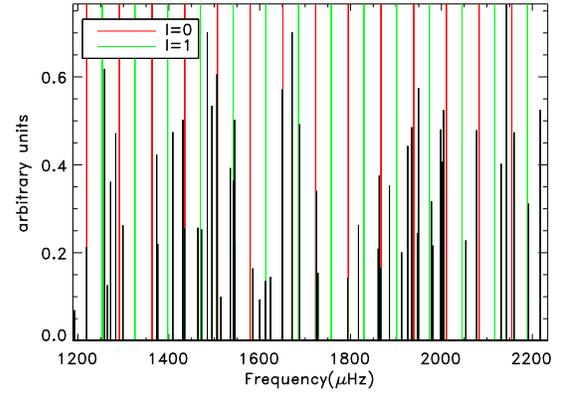}}

\vspace{0cm}
      \caption[]{Example of the CLEAN-est spectrum computed with the local weights (Foster, 1995). The color lines in the figure correspond to the frequencies from the asymptotic relation with a mean large splitting of 74\,$\mu$Hz.
}
         \label{Fig_sfsbvir}
   \end{figure}

In Fig.~\ref{Fig_sfsbvir}, we show an example of the CLEAN-est spectrum  computed with the local weights \citep{Foster95}. The color lines in the figure correspond to the frequencies from the asymptotic relation with a mean large splitting of 74\,$\mu$Hz. Some peaks correspond probably to the sidelobes (at $\pm\,11.6\mu$Hz) chosen by CLEAN-est procedure instead of the central peaks in power spectrum.

% Fig_sfsbvir.tex

%\resizebox{4.6in}{!}{\includegraphics{combvir85.eps}}

By adjusting the weights for two instruments we expect to obtain better determination of the large splitting $\Delta\nu_{\rm 0 }$. The frequency spacing of 74\,$\mu$Hz, as determined from present comb analysis, is consistent with theoretical predictions for $\beta$\,Vir and will be compared with the
results from evolution models \citep{Fernandes04}.

%-----------------------------------------------------------------------------

\section{\textsc{Conclusion}}

The observations of $\beta$\,Vir show the presence of an excess power around 1.7\,mHz, which is likely to be due to solar-like $p$-modes.
 It was detected independently with two instruments (FEROS and ELODIE) on two different sites and using different calibration methods (simultaneous thorium and channelled spectrum from Fabry-Perot interferometer).
The estimation of the upper limit for the detection of the oscillation amplitude is about 45\,cm/s, in agreement with expectations for this star. The identification of the $p$-modes frequencies
is in progress and will be presented in the next paper. 

\footnotesize{\textit{Acknowledgements.} We are grateful to OHP and FEROS staff for the on site support of the observations.
M.M. was funded by ESA/SSD under contract C15541/01.}

%\section*{Acknowledgements}
%{\footnotesize
%We acknowledge INSU/PNPS for the financial support.
%}

%-----------------------------------------------------------------------------
%\vspace{1cm}

\end{document}